\documentclass{article}

\usepackage{arxiv}
\usepackage{amsmath}
\usepackage[utf8]{inputenc} 
\usepackage[T1]{fontenc}    
\usepackage{hyperref}       
\usepackage{url}            
\usepackage{booktabs}       
\usepackage{amsfonts}       
\usepackage{nicefrac}       
\usepackage{microtype}      
\usepackage{lipsum}		
\usepackage{graphicx}

\usepackage{doi}
\usepackage[
    backend=biber,natbib,
    style=numeric
]{biblatex}
\title{GraphCNNpred: A stock market indices prediction using a Graph based deep learning system}

\author{ 
	{Yuhui Jin} \\
	Department of Mathematics\\
	California Institute of Technology\\
	Pasadena, California\\
	\texttt{yjjin@caltech.edu} \\
}

\renewcommand{\shorttitle}{Graph-CNNpred}

\hypersetup{
pdftitle={Graph-CNNpred:  A stock market indices prediction using a Graph based deep learning system},
pdfsubject={q-bio.NC, q-bio.QM},
pdfauthor={Yuhui Jin},
pdfkeywords={First keyword, Second keyword, More},
}

\begin{document}
\maketitle

\begin{abstract}
The application of deep learning techniques for predicting stock market prices is a prominent and widely researched topic in the field of data science. To effectively predict market trends, it is essential to utilize a diversified dataset. In this paper, we give a graph neural network based convolutional neural network (CNN) model, that can be applied on diverse source of data, in the attempt to extract features to predict the trends of indices of \text{S}\&\text{P} 500, NASDAQ, DJI, NYSE, and RUSSEL. The experiments show that the associated models improve the performance of prediction in all indices over the baseline algorithms by about $4\% \text{ to } 15\%$, in terms of F-measure. A trading simulation is generated from predictions and gained a Sharpe ratio of over 3.
\end{abstract}


\section{Introduction}
The financial market plays a crucial role in the development of global economies. The stock market experiences enormous trading volumes daily. The efficient market hypothesis suggests that all market information should be promptly reflected in an ideally competitive stock market. Consequently, analyzing stock behaviors and predicting future trends has been a long-standing research topic in the field of finance. However, the fluctuation of stock prices is influenced by numerous factors, including the global economic environment, company financial reports, news and events that may or may not be directly related to finance, and financial experts' speculations. As machine learning, particularly deep learning, techniques have flourished in the past 10 years, predicting stock behavior has become an active topic in data science. In this paper, we propose a graph-based model for predicting stock index trends using diversified data sources.

This paper is organized as follows: In Section 2, related works are presented. In Section 3, we briefly introduce the dataset. In Section 4, our proposed method is detailed, starting with an introduction to various utilized background information as well as experimental setups. In Sections 5 and 6, we present the results and discussion of our model, along with a trading simulation. In Section 7, the conclusion is presented.

\section{Related Works}
Predicting the price trends or movements of individual stocks can be challenging due to various factors such as company operations, public opinion, and other unknown underlying factors. Stock market indices provide an alternative, presumably more reliable, approach for understanding the overall trend of the general stock market as they are weighted aggregations from individual companies. There are roughly two types of models (not necessarily mutually exclusive): those that predict individual stock prices and those that predict indices. \cite{meta_survey} offers an extensive survey of deep learning techniques applied to stock market prediction. In this section, we will only present some works specifically utilizing CNN or graph methods.

\subsection{Convolutional Neural Networks on Stock Predictions }

Although CNN models were originally developed for computer vision tasks, they can be adapted for stock prediction by processing time series data and extracting one-dimensional data from the input matrix. Typically, CNNs use one-dimensional filters to slide over the time series data with a stride determined by the data size.

\cite{cnn_event} improved upon the event embedding method \cite{cnn_even_pre} by utilizing a CNN to train on input history event data. The pooling layer effectively extracts representative event history features.

\cite{ucnnpred} (U-CNN pred), originating from \cite{CNNpred}, was trained using a layerwise approach, where the sub-CNN layers were pretrained sequentially until the proposed model structure was completed. This model showed good results and was effective due to its simple architecture and shallow layers, which reduced the curse of dimensionality, as there were fewer weights to be learned.

To fully utilize the information contained within predetermined industrial relations, some models have integrated knowledge graphs with CNNs. The Knowledge-Driven Temporal Convolutional Network (KDTCN) proposed by \cite{cnngraph} utilizes Open IE to extract events related to the knowledge graph and make stock predictions. To avoid common data leakage from one-dimensional convolutional layers, KDTCN employs causal convolution, which only uses information from the current and previous time steps in the previous layer. KDTCN is effective in explaining abrupt price changes by extracting significant features in the price time series \cite{cnngraph}.
\subsection{Graph Neural Networks on Stock Predictions}
There have been interesting works (\cite{temporal},\cite{hats},\cite{exp_temp}) that incorporate knowledge graphs into their stock prediction models for individual companies by combining graph neural networks and sequential models such as Recurrent Neural Networks (RNN). The general recipe for such models is as follows:
1. Historical stock data are fed into a time series layer as features to output node embeddings for each stock: previous day’s closing price and moving averages \cite{temporal}, price change rates \cite{hats}.
2. The adjacency matrix for the graph is a pre-determined graph created through open knowledge graph sources to connect companies through industry, subsidiary, people, and product relations.
3. The node embeddings and graph information are fed into a graph neural network layer to update the node embeddings: rolling window analysis is performed in \cite{exp_temp}.
4. The updated embeddings are fed into a fully connected layer for predictions.
All these works rely on the original or some derived version of GAT and GCN.

\cite{overnight} proposed an LSTM Relational Graph Convolutional Network (LSTM-RGCN) model that handles correlations among stocks, where correlations are calculated based on historical market data. The LSTM mechanism is added to the RGCN layers to alleviate the over-smoothing problem.

\cite{mg-conv} employed the use of GCN to analyze the correlation of indicators in stock trend prediction. MG-Conv is based on a multi-Graph Convolutional Neural Network and utilizes static graphs between indices that are constructed using constituent stock data. Moreover, they created dynamic graphs based on trend correlations between indices with different portfolio strategies and defined multi-graph convolution operations based on both graphs.
\section{The Dataset}
The dataset we use in this paper is originally from \cite{CNNpred}, so we will only provide a brief outline here. We have 82 daily features for each market. These can be categorized into eight different groups: primitive features, technical indicators, economic data, world stock market indices, the exchange rate of the US dollar to other currencies, commodities, data from big companies in the US market, and future contracts. We emphasize that only primitive features and technical indicators vary between different indices.

\section{Methodology}
\subsection{Background}
In this subsection, we will provide an overview of two categories of neural networks in GraphCNNPred: graph neural network (GNN) and CNNpred.
\subsubsection{Convolutional Neural Network and CNNpred}
Convolutional Neural Networks (CNNs) were introduced in 1995 by \cite{CNN} and have since been applied for feature selection and market predictions in the finance industry. Our model, GraphCNNPred, is a derived and generalized version of CNNpred, a CNN-based framework for stock market prediction \cite{CNNpred}. CNNpred comprises of two variations known as 2D-CNNpred and 3D-CNNpred. We will explain its framework in four main steps: representation of input data, daily feature extraction, durational feature extraction, and final prediction.

Representation of input data:
CNNpred gathers information from various markets to predict their future trends. The objective is to create a universal model capable of mapping a market's historical data to its future fluctuations. By "universal model," we mean a model that applies across multiple markets. Thus, the model must be trained using data samples from different markets to learn this mapping function accurately.

Daily feature extraction:
Each day, the historical data consists of closing prices of particular assets, closing prices of derivatives  and other technical factors. The design of the first layer in CNNpred is based on this observation. In both variations of CNNpred (2D-CNNpred, 3D-CNNpred), the initial step involves a convolutional layer tasked with aggregating daily variables into a  higher-level feature that represent each day in the historical record.

Durational feature extraction:
Understanding the future behavior of markets often requires extracting information directly from the temporal evolution of market behavior. This involves analyzing patterns and trends over consecutive days of data to extract high-level features that reflect the market's evolving behavior.

Final prediction:
In the last step, the features generated in the preceding layers are transformed into a one-dimensional vector through a flattening operation. This vector is then fed into a fully connected layer, which maps these features to a prediction.

\begin{figure}
    \centering
    \includegraphics[width=0.9\linewidth]{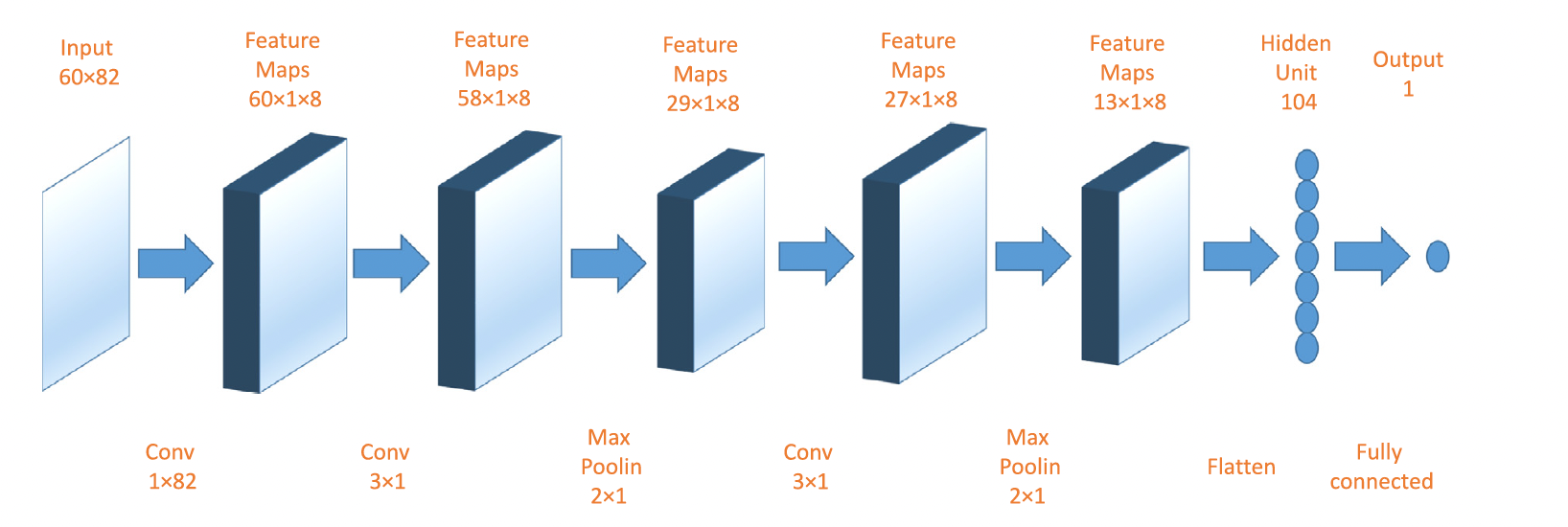}
    \caption{CNN-pred visualization}
    \label{fig:enter-label}
\end{figure}

\subsubsection{Graph Neural Network}

Graph-based structures have been successful in expressing inductive bias for complex interconnected systems, leading to applications in biology, physics, chemistry, and recommendation systems. Typical tasks include node prediction, graph embedding, and graph-level classification. In recent years, Graph Neural Networks (GNNs) have been successfully applied to all these different learning problems. GNNs leverage the topological structure of graphs to encode information about entities such as nodes, edges, or hypergraphs.

At the heart of the GNN model is the notion of \textit{message passing}. In each layer of the network, each node feature is updated according to a "message" composed of features from adjacent nodes. For a graph $G$ with vertex set  $V(G)$ and node features ${h_v}_{v\in V(G)}$, the node features at step $n$ are updated via:
\[
h_v^{n+1} = \mathrm{AGG}(h_v^{n}, \mathrm{MSG}(\{h^{n}_u\}_{u\in \mathcal{N}_v}))
\]
where  $\mathrm{MSG}$ and $\mathrm{AGG}$ can generally be learned. Various choices and constraints of these functions give rise to different neural network architectures. For graph classification tasks, the node features of the final layer \(\{h_v^{N}\}\) are pooled together. Thus, in this approach, we diffuse messages through the graph and then pool them together.

We elaborate on two classical GNNs that constitute the graph part of our model: Graph Convolution Network (GCN) \cite{GCN} and Graph Attention Network (GAT) \cite{GAT}. An extra assumption is imposed here for the rest of this section: no edge information is allowed. There are two major steps for one GCN layer: message passing and combination. We denote $u,v$as nodes, $x_u, x_v$ as node features, $N_v$ as the neighborhood of $v$, and $d_v$ as the degree of $v$. For message passing,
\[p_v = \text{AGG}(\{x_u \mid u \in N_v\}) = \sum_{u \in N_v} \frac{x_v}{\sqrt{d_v d_u}}.\]
For combination,
\[h_v = \text{ReLU}(W(\frac{x_v}{d_v} + p_v)),\]
where \(W\) is a trainable weight matrix. 

One natural question to ask for GCN is: Can we implicitly learn the aggregation weight, contrary to assigning each node a weight $1/\sqrt{d_vd_u}$? GAT \cite{GAT} provides a solution: A shared linear transformation, parameterized by a weight matrix $W$, is applied to all nodes, and a shared attention mechanism that computes attention coefficients will indicate the importance of node $u$ to node $v$. Thus, GAT is constructed by replacing attention coefficients with the original GCN node weight.

\subsection{Our model}
In \cite{CNNpred}, both $\text{2D-CNNpred}$ and $\text{3D-CNNpred}$ are structured around three main stages: 1. extracting daily features, 2. handling duration data, and 3. making final predictions. Traditionally, daily features are extracted using a linear transformation that reduces the 82-dimensional data to a single dimension. However, due to the intricate and implicit nature of financial data and its features, we propose an alternative approach for daily feature extraction. While the steps for handling duration data align with CNNpred for its simplicity and low variance construction compared to traditional time series models like RNN or LSTM, we introduce flexibility by removing the constraint of the order in the CNNpred model.

This flexibility allows our model to be configured in three possible layouts:
1. Starting with a CNN for duration data and concluding with daily feature extraction using graphs.
2. Beginning with a small CNN to partially extract duration data, followed by daily feature extraction using graphs, and ending with another small CNN to handle all duration data.
3. Initiating with daily feature extraction using graphs and concluding with a CNN for duration data.

Constructing graphs and utilizing graph layers is more intricate than employing a single CNN layer, requiring more parameter tuning and trainable coefficients, especially with the self-attention mechanism of GAT. To manage this complexity, we balance the model with straightforward CNN time series layers, which help mitigate the curse of dimensionality.

\subsubsection{Construction of graph}
We compute correlation maps for all 82 features using only the training data. Each feature is treated as a node in the graph, and an edge is formed between two features if their correlation exceeds a fixed threshold — in this study, we set this threshold to 0.7. Graph Neural Networks (GNNs) typically leverage such topological structures to encode graph information, enabling message passing from nodes to their neighbors in each layer. Therefore, nodes connected by an edge share information during the message passing process. Conversely, features with low or no correlation are considered independent and remain isolated; after the graph layer, their information depends solely on global and local contexts.
In our model setup, to extract daily features — i.e., to capture information from all features on a given day — we incorporate the information of highly correlated features (i.e., their neighborhood) using GCN or GAT.

\subsubsection{GAT-CNNpred}

Representation of input data: As previously mentioned, the input data structure mirrors that of 2D-CNNpred, forming a two-dimensional matrix. The matrix size is determined by the number of variables representing each day and the number of days included in the historical data used for prediction. If the prediction input comprises $d$ days, each represented by $f$ variables, then the size of the input tensor will be $d\times f$.

Daily feature extraction: In CNNpred, daily feature extraction involves using $1 \times $ number of initial variable filters. Each filter encompasses all daily variables and amalgamates them into a single higher-level feature representation. In contrast, GAT-CNNpred employs a Graph Attention Network (GAT). As discussed in the Construction of Graphs subsection, each feature is treated as a node in the graph, and edges are formed based on correlations exceeding a predefined threshold. A single layer of GAT enables each feature node to gather information from other correlated features within the network. Consequently, the output of this layer encapsulates modified data for each node, integrating influences from its neighboring nodes.

Durational Feature Extraction: In CNNpred, durational feature extraction utilizes convolutional layers with $3\times 1$ filters to aggregate features from different days into higher-level representations, thereby capturing information over specific durations. Each $3\times 1$filter spans three consecutive days, a design influenced by well-known candlestick patterns such as Three Line Strike and Three Black Crows, which analyze meaningful patterns over short time frames \cite{candlestick}. Following the convolutional layers, an activation layer and a max pooling layer are applied. 
To capture information over longer time intervals and create more complex features, 2D-CNNpred extends this approach with additional convolutional layers using $3\times 1$ filters, followed by pooling layers similar to the initial setup. 

In our proposed generalized approach to durational feature extraction inspired by CNNpred, rather than strictly sequential layers, we consider these convolutions as independent sets. This allows flexibility in how these layers interact, focusing on extracting features from distinct time periods without strict ordering constraints. These intersections with graph layers will be further detailed in the sample configuration. By integrating this approach with daily feature extraction, the model effectively extracts and combines data features across different dates and variables, enhancing its capability to capture diverse temporal and feature-based patterns.

Final Prediction: After passing through the last pooling layer, the features are flattened into a final feature vector. This vector is then processed through a fully connected layer with a sigmoid activation function to produce the final prediction. Each prediction consists of five numbers for each data point, representing the probabilities of price increases for the next day across five indices.
This probabilistic output allows for a nuanced approach to trading decisions, where more significant investments can be allocated to stocks with higher predicted probabilities of price increases, and vice versa.
In Section 5, the model's output is discretized to either 0 or 1, depending on which is closer to the predicted probability. In Section 6, $(0,1,2)$labeling system is employed, resulting in a prediction vector of 15 dimensions where the output is discretized into one of three classes: 0, 1, or 2. Explicitly, the position of maximum among the three scores toward one index is the output.  The detail explanation about this label is in Section 4.3.

A sample configuration of GAT-CNNpred:As previously described, each input for prediction comprises 60 days, each represented by 138 variables (details explained in the Data Preparation subsection). Therefore, the input to GAT-CNNpred is a matrix of size 60 by 138. Below, we outline three submodels:
\begin{itemize}
    \item  GAT-CNN:The first two graph layers (GAT) have output channels of 30 and 10, respectively. These are followed by a pooling layer that summarizes graph information into a single-node vector shape. Then, there are two convolutional layers with 8 filters of size  $5\times 1$ each, followed by a ReLU activation layer and a $2\times 1$ max-pooling layer.

    \item  CNN-GAT:The first two convolutional layers consist of 8 filters of size $5\times 1$ each, followed by a ReLU activation layer and a $2\times 1$ max-pooling layer. Subsequently, two graph layers (GAT) with output channels of 30 and 10, respectively, are employed. These are then followed by a pooling layer that summarizes graph information into a single-node vector shape.

    \item  CNN-GAT-CNN:The model starts with a convolutional layer featuring 8 filters of size $5\times 1$, followed by a ReLU activation layer and a $2\times 1$ max-pooling layer. Next, two graph layers (GAT) with output channels of 20 and 10, respectively, are utilized. Following this is a pooling layer that summarizes graph information into a single-node vector shape. Finally, there is another convolutional layer with 8 filters of size $5\times 1$  each, followed by a ReLU activation layer and a $2\times 1$  max-pooling layer.
    
\end{itemize}
The final flattened feature vector is fed into a fully connected layer to produce the final output. Unlike models predicting a single index trend, our model generates predictions for 5 indices simultaneously. The choice of pooling layer for the graph will be detailed in the subsection on pooling. In summary, three pooling methods are considered: mean, max, and fully-connected.

\subsubsection{GCN-CNNpred}
This model is akin to GCN-CNNpred. However, it's important to note that GCN uses fixed aggregation weights, while GAT incorporates a shared attention mechanism to compute attention coefficients. Achieving similar model performance with both graph models typically requires more layers in GCN compared to GAT. To maintain a balanced model complexity, the output channels in GCN decrease more rapidly than in GAT. A common configuration for GCN-CNNpred involves 6 layers of GCN with output channels set as $10, 7, 2, 3, 5, 5$.

\subsubsection{Pooling for Graphs}
Given a graph with $N$ nodes, $F$ features, and a feature matrix $X$ (with $N$ rows and $F$ columns), pooling condenses this graph into a single node in a single step. We introduce three pooling methods for graphs:
\begin{itemize}
    \item Mean Pooling: Computes the feature-wise mean across the node dimension to produce the output of the pooling layer.
   
    \item Max Pooling: Computes the feature-wise maximum across the node dimension to produce the output of the pooling layer.
   
    \item Fully Connected Layer Pooling: Utilizes a fully connected layer with trainable variables of size $F$ to aggregate information.
\end{itemize}

According to \cite{pooling}, each pooling method has distinct properties and applications in graph neural networks:
\begin{itemize}

    \item Mean Pooling: Performs well when statistical and distributional information in the graph is crucial, rather than the exact node structure. It is effective when node features are diverse and do not repeat frequently.

    \item Max Pooling: Identifies representative elements or the "skeleton" of a graph, making it suitable for tasks where identifying key features is more important than preserving exact structural details. It is robust to noise and outliers.

    \item Fully Connected Layer Pooling: Acts as a decoder in graph neural networks, offering flexibility in aggregating features based on learned parameters.
\end{itemize}
In our model, we incorporate all three pooling layers and select the optimal pooling method based on performance metrics specific to our task.

\subsection{Data preparation}

In this paper, we focus on short-term price trend prediction using daily closing prices from the original dataset. To label the data, we calculate the daily return of day $t$ as $\frac{\text{Close}_{t+1}}{\text{Close}_{t}} - 1$, where $\text{Close}_{t}$ is the closing price on day $r$. We propose two labeling systems. The first system categorizes returns as follows: if the return is greater than 0, it is labeled as 1; otherwise, it is labeled as 0. We call it (0,1) labeling. However, distinguishing between returns like $-0.0001$ and $0.0001$ is challenging for the model. In real trading scenarios, predicting a slight negative return should not be considered as detrimental as predicting a large drawdown. Therefore, we introduce a second labeling system. For all training data, we label returns based on thresholds derived from the 35th and 65th percentiles of returns: label 2 for returns above the high threshold, label 0 for returns below the low threshold, and label 1 otherwise. We call it (0,1,2) labeling. This labeling approach allows us to consider multi-class predictions for short-term returns, which are crucial in financial markets.
Additionally, we extend the concept of one-day returns to n-day returns, where n ranges from 1 to 10, reflecting the short-term nature of the predictions. 

The dataset covers the period from January 2010 to November 2017, and we split it into training ($65\%$), validation ($15\%$), and test ($20\%$) sets.

Given the diverse ranges of data, normalization is essential. Our model heavily relies on graph neural networks with various pooling techniques. Non-standardized data can lead to issues such as extensive gradient vanishing and slow learning curves. Therefore, we apply standard normalization to all data points: $\tilde{X}_{i,j}=(X_{i,j}-\bar{X}_i)/\text{std}_i$, where $\bar{X}_i),\text{std}_i$ are computed from training data. 

In Section \ref{sec:sim}, all data are integrated into the combined simulation part. Notably, the feature dimension used is 138, rather than 82, as 3D-CNNpred achieved slightly better performance in terms of F-measure. This improvement is likely due to the comprehensive utilization of all available data features.

\subsection{Evaluation}

Evaluation metrics are crucial for model selection, especially in the context of imbalanced datasets. Typically, for such datasets, we rely on the Macro-Averaged F-measure (mean of F-measures) for each of the two classes \cite{f1mean}. In Section 5, we will use binary (0, 1) labeling and F-measure as the metric. In Section 6, where a 0,1,2 labeling system is employed, we will directly use the Sharpe value and CEQ from trading simulations as a criterion.

\subsection{Parameters of the Network}

In this work, PyTorch was utilized for implementing CNNs, while Torch-Geometric was employed for graph neural networks. The threshold for forming graph edges was set at a correlation of 0.7. All layers used the ReLU activation function, except for the final layer which used Sigmoid. The configurations described in the GAT-CNNpred subsection and their counterparts in GCN-CNNpred are considered as candidates in the following sections. Adam optimizer with a batch size of 32 was used for training the networks.

\section{Conclusions and Discussions}

\subsection{Result}

We compared all proposed Graph-CNNpred models with algorithms used in CNNpred and pure graph techniques. All baseline algorithms were tested using the settings reported in their original papers. Recall that 2D-CNNpred and 3D-CNNpred from \cite{cnnpred} statistically outperformed baseline algorithms such as PCA-ANN, technical indicators, and CNN-cor. Therefore, CNNpred stands out as an impressive model and serves as a strong baseline.
Additionally, we included graph techniques (GAT and GCN) as additional baseline algorithms to compare the performance of our graph-inspired models with their original graph techniques. For both graph techniques, initial data was fed into graphs constructed based on feature correlations, followed by shallow ANN for prediction.

To ensure fairness, each baseline algorithm was tested multiple times under the same conditions, and their average F-measures were compared. Sections 5 and 6 will present results from two different train/test splits: the current split ($65\%-15\%-20\%$, starting from $04/21/16$) yielding a Sharpe ratio of 0.114 or annual Sharpe of 1.8 for an always-long strategy. However, historical data from 2013 to 2023 shows that an average Annualized Sharpe of a always long strategy for indices sits 0.84. Thus, a proposed new split ($42\%-8\%-50\%$, starting from $04/21/14$) resulting in a Sharpe ratio of 0.043 or annual Sharpe of 0.7 for the same strategy.

Table 1 summarizes the average results for baseline algorithms and our suggested models across the S$\&$P 500 index, Dow Jones Industrial Average, NASDAQ Composite, NYSE Composite, and RUSSELL 2000 historical data in terms of F-measure. The difference between baseline algorithms and GCN-CNNpred and GAT-CNNpred is statistically significant. The best-performing algorithms across different indices are also reported in Table 2. Tables 3 and 4 show the mean and best performances when test dates start from 2014.

\begin{table}
	\caption{Best F-measure of various algorithms}
	\centering
	\begin{tabular}{llllll}
		\toprule
		
		Strategy     & S\&P 500     &DJI &NASDAQ & NYSE & RUSSEL \\
		\midrule

            2D-CNNpred& 0.5408  & 0.5562 &0.5521 & 0.5472 & 0.5463  \\
            3D-CNNpred & 0.5532 &0.5612&0.5576 &0.5592 &0.5787 \\
            GAT &  0.5740  &0.5495 & 0.5523 &0.5544  &0.5702  \\
            GCN & 0.5310  & 0.5791 &0.5590 &  0.5405 & 0.5574 \\
            GAT-CNNpred  & 0.5819  & 0.5861 &0.5775 & 0.5907 & 0.5815 \\           
            GCN-CNNpred  & 0.5866  & 0.5885 &0.5741 & 0.5767  &0.5954  \\

		\bottomrule
	\end{tabular}
	\label{best_f_16}

\end{table}

\begin{table}
	\caption{Mean F-measure of various algorithms}
	\centering
	\begin{tabular}{llllll}
		\toprule
		
		Strategy     & S\&P 500     &DJI &NASDAQ & NYSE & RUSSEL \\
		\midrule

            2D-CNNpred& 0.4914  & 0.4975 &0.4944 & 0.4885 & 0.5002  \\
            3D-CNNpred & 0.4837 &0.4979&0.4931 &0.4751 &0.4846 \\
            GAT &  0.4673  &0.4756 & 0.4619 &0.4782  &0.5034  \\
            GCN & 0.4458  & 0.4546 &0.4569 &  0.4590 & 0.5008 \\
            GAT-CNNpred  & 0.5121  & 0.5334 &0.5098 & 0.5223 & 0.5191 \\           
            GCN-CNNpred  & 0.5299  & 0.5237 &0.5158 & 0.5274  &0.5312  \\

		\bottomrule
	\end{tabular}
	\label{tab:table_sharpe}

\end{table}

\begin{table}
	\caption{Best F-measure of various algorithms from 2014}
	\centering
	\begin{tabular}{llllll}
		\toprule
		
		Strategy     & S\&P 500     &DJI &NASDAQ & NYSE & RUSSEL \\
		\midrule

            2D-CNNpred& 0.5179 & 0.5031 &0.5249 & 0.5159 & 0.5224  \\
            3D-CNNpred & 0.5030 &0.5099&0.5435 &0.4986 &0.5200 \\
            GAT & 0.5230 &0.5329 &0.5336 & 0.5386  & 0.5325\\
            GCN & 0.5051 &0.5206 &0.5477  &0.5275& 0.5504 \\
            GAT-CNNpred  & 0.5362  & 0.5457 &0.5516 & 0.5517 & 0.5692 \\           
            GCN-CNNpred  & 0.5359  & 0.5430 &0.5628 & 0.5642  &0.5443  \\

		\bottomrule
	\end{tabular}
	\label{tab:table_sharpe}

\end{table}

\begin{table}
	\caption{Mean F-measure of various algorithms from 2014}
	\centering
	\begin{tabular}{llllll}
		\toprule
		
		Strategy     & S\&P 500     &DJI &NASDAQ & NYSE & RUSSEL \\
		\midrule

            2D-CNNpred& 0.4674  & 0.4663 &0.4620 & 0.4690 & 0.4704  \\
            3D-CNNpred & 0.4570 &0.4754&0.4819 &0.4769 &0.4855 \\
            GAT &  0.4729  &0.4723 & 0.4775 &0.4532  &0.4993  \\
            GCN & 0.4441  & 0.4599 &0.4693 & 0.4411 & 0.4620 \\
            GAT-CNNpred  & 0.5194  & 0.5152 &0.5199 & 0.4918 & 0.5223 \\           
            GCN-CNNpred  & 0.4983  & 0.5210 &0.4998 & 0.5093  &0.5054  \\

		\bottomrule
	\end{tabular}
	\label{tab:table_sharpe}

\end{table}

\subsection{Discussion}

The results clearly indicate that both GAT-CNNpred and GCN-CNNpred statistically outperform baseline algorithms, including CNNpred and traditional graph techniques, as evidenced by the difference in F-measure.

Recall that CNNpred extracts daily features using a simple fully connected layer, whereas Graph-CNNpred facilitates message passing among features. CNNpred treats the extraction of information from daily and durational features as orthogonal tasks, making it challenging to discern complex patterns across short time intervals and correlated features simultaneously. In contrast, GCN and GAT do not process data in a time series manner but excel in identifying daily feature characteristics.

In GCN-CNNpred, the model leverages CNN layers for capturing durational information and GCN layers for aggregating highly correlated data over specific time periods, extracting sophisticated features from this two-dimensional data concurrently. GAT-CNNpred incorporates an attention mechanism that efficiently gathers correlated features, theoretically enhancing training efficiency.

Our hybrid framework of integrating graph techniques with CNN offers two distinct advantages over traditional baseline algorithms: First, it respects the time series nature of the dataset without overly complicating prediction procedures. Second, it utilizes a comprehensive set of features that contain valuable information for stock prediction, overcoming the orthogonality constraints of time and feature dimensions. A well-designed CNN and graph technique combination enhances results and ensures robust and efficient learning procedures.

\section{Trading Strategies and Simulations}
\label{sec:sim}
In this section, we focus on trading strategies derived from all Graph-CNN models discussed earlier. Theoretically, models with high prediction accuracy should yield strong trading performance. Our models notably outperform many contemporary market prediction systems. A key question here is whether these graph-based models can generate effective trading strategies and performance.

Recall from the labeling subsection, we introduced two labeling systems, the second being a three-category labeling strategy. Specifically, the second labeling categorizes daily stock returns as "up" if significantly positive, "down" if significantly negative, and "unchanged" if close to zero. During implementation, these categories are encoded using one-hot encoding.

This labeling directly informs our trading strategy: if the highest prediction dimension on day t indicates an upward movement (dimension 1), we take a long position. Similarly, if it indicates a downward movement (dimension 2), we take a short position. If the prediction suggests minimal change (dimension 3), we close any existing positions or refrain from trading. We evaluate performance using the Sharpe ratio:

\[\text{Sharpe}=\frac{\sum_{t=1}^n r_t/n}{\text{std}(r_1,...,r_n)},\] where $r_t$ denotes the profit and loss (PnL) on day t. In financial trading, the annualized Sharpe ratio, adjusted by multiplying by $\sqrt{252}$, is commonly used. Transaction costs are excluded in this analysis.Another performance measure is the certainty-equivalent (CEQ) return \cite{ceq}: \[\text{CEQ}= \sum_{t=1}^n r_t/n-\frac{\gamma {\text{std}(r_1,...,r_n))^2}}{2},\] where $\gamma$ represents the risk aversion coefficient, set to 1 in this study.

The following is a list of model we used in trading strategy.
\begin{itemize}
    \item GAT: Pure GAT strategy without any CNN layers, with 012 labeling.
    \item GCN: Pure GCN strategy without any CNN layers, with 012 labeling.
    \item GAT-CNNpred: A strategy combined with a two layers of GAT and convolution layers for duration information, with 012 labeling. Possible combination includes different style of graph pooling, GAT then CNN, CNN then GAT, CNN then GAT then CNN. 
    \item GCN-CNNpred: A strategy combined with a two layers of GCN and convolution layers for duration information, with 012 labeling. Possible combination includes different style of graph pooling, GCN then CNN, CNN then GCN, CNN then GCN then CNN. 
\end{itemize}
In Table\ref{tab:table_sharpe}, Combination is a column indicating that we can long or short one 1 unit of any five indices.
During the simulation, one key situation that rings a bell is the always long strategy combination Sharpe. From the subsection above regarding result, there are two set of train-test split: $65\%-15\%-20\%$ and  $42\%-8\%-50\%$.
\begin{table}
	\caption{Sharpe ratio of various algorithms}
	\centering
	\begin{tabular}{lllllll}
		\toprule
		
		Strategy     & S\&P 500     &DJI &NASDAQ & NYSE & RUSSEL & Combination \\
		\midrule
		Always long & 0.1056  & 0.1472 &0.1347 & 0.0753 & 0.0739  &0.1143 \\
            2D-CNNpred& 0.1422  & 0.1703 &0.1163 & 0.1039 & 0.08039 &0.1302  \\
            3D-CNNpred & 0.1413 &0.1344&0.1642 &0.0830 &0.0910  & 0.1458\\
            GAT &  0.1981  &0.1980 & 0.1742 &0.1597 & 0.1679 &0.1822  \\
            GCN & 0.1732  & 0.1801 &0.1794 &  0.1345 & 0.1585 &0.1684 \\
            GAT-CNNpred  & 0.2017  & 0.2191 &0.2301 & 0.1882 & 0.1614 &0.2116 \\           
            GCN-CNNpred  & 0.1807  & 0.1805 &0.2258 & 0.1848 & 0.1401 &0.2123  \\
             
		\bottomrule
	\end{tabular}
	\label{tab:table_sharpe}
\end{table}

\begin{table}
	\caption{CEQ return of various algorithms}
	\centering
	\begin{tabular}{lllllll}
		\toprule
		
		Strategy     & S\&P 500     &DJI &NASDAQ & NYSE & RUSSEL & Combination \\
		\midrule
		Always long & 0.0004946  & 0.0006717 &0.0008389 & 0.0003598 & 0.0005738  &0.002710 \\
            2D-CNNpred& 0.0006681  & 0.000776 &0.0007234 & 0.0004988 & 0.0006265  &0.002904 \\
            3D-CNNpred& 0.000664  & 0.0006129 &0.001023 & 0.0003974 & 0.0007127  & 0.002956 \\
            GAT &0.0006918  & 0.0008383 &0.001205& 0.0006943 & 0.0009215 &0.003147  \\
            GCN & 0.0005877  &0.0008379 &0.001205 & 0.0006485 & 0.0008953 &0.003201\\
            GAT-CNNpred &0.0009047 &0.0009195 &0.001423 &0.001095 &0.001279&0.003535  \\ 
            GCN-CNNpred  & 0.0008480 & 0.0008608 &0.001388 & 0.0008230 & 0.001188 &0.003388  \\

		\bottomrule
	\end{tabular}
	\label{tab:table_ceq}
\end{table}

\begin{table}
	\caption{Sharpe ratio of various algorithms since 2014}
	\centering
	\begin{tabular}{lllllll}
		\toprule
		
		Strategy     & S\&P 500     &DJI &NASDAQ & NYSE & RUSSEL & Combination \\
		\midrule

		Always long & 0.04350  & 0.05173 &0.05864 & 0.02016 & 0.03295  &0.04283 \\
            2D-CNNpred& 0.03249  & 0.06591 &0.05792 & 0.06927 & 0.07392 &0.06943  \\
            3D-CNNpred& 0.07093  & 0.06001 &0.08066 & 0.05491 & 0.06043 &  0.06929  \\
            GAT &  0.08051  &0.08901 & 0.08687 &0.07584 &0.08219 &0.07792  \\
            GCN & 0.09596  & 0.08453 &0.09457 &  0.09191 & 0.07641 &0.08809 \\
            GAT-CNNpred  &  0.1013  & 0.1118 &0.1974 & 0.09564 & 0.1166 & 0.1311 \\           
            GCN-CNNpred  & 0.1030  & 0.09181 &0.1073 & 0.11267 & 0.11148&0.1126  \\

		\bottomrule
	\end{tabular}
	\label{tab:table_sharpe}
\end{table}

\begin{table}
	\caption{CEQ return of various algorithms since 2014}
	\centering
	\begin{tabular}{lllllll}
		\toprule

		Strategy     & S\&P 500     &DJI &NASDAQ & NYSE & RUSSEL & Combination \\
		\midrule

		Always long & 0.0003110  & 0.0003683 &0.0004957& 0.0001285 & 0.0002826  &0.0009341 \\

            2D-CNNpred& 0.0004279  & 0.0005432 &0.0004458 & 0.0005043 & 0.0005261 &0.001175 \\
            3D-CNNpred& 0.0003593  & 0.0006948 &0.0005395 & 0.0006311 & 0.0005062 &0.001355\\
            GAT & 0.0005321  & 0.0005414& 0.0006943 &0.0006642 & 0.0007338 &0.001437  \\
            GCN & 0.0005643  &0.0006526&0.0008241 &0.0007352 & 0.0006844 &0.001576\\
            GAT-CNNpred &0.0006412 &0.0008134 &0.001136 &0.0006403 &0.0006399& 0.001654  \\ 
            GCN-CNNpred  &0.0007228  &0.0006576 & 0.0009385 &0.0007246 & 0.001045 &0.001887  \\

		\bottomrule
	\end{tabular}
	\label{tab:table_ceq}
\end{table}

\subsection{Discussions}

Graph-CNNpred was tested as part of a stock market trading system to assess its impact on trading performance using standard evaluation measures for trading strategies. While it's intuitive that a model with accurate market predictions can potentially enhance trading performance, the crucial question remains: how much can a robust predictive model actually contribute to profits in a simulated trading environment?

Our experiments demonstrate that employing CNNpred predictions for trading strategies yields favorable results in terms of Sharpe ratio and CEQ return measures across most tested indices and combinations thereof. As a baseline, the "always long" strategy (buy at day 1 and always hold) serves as a passive approach. Additionally, strategies derived from 2D-CNNpred, 3D-CNNpred, GAT, and GCN were evaluated. These findings highlight that the Graph-CNNpred framework shows promise as a viable candidate for integration into real-world trading systems as a prediction module.

\section{Conclusion}

Predicting financial market prices is a challenging task due to their nonlinear and sophisticated nature. Improved predictions hinge on high-quality data variables. In this paper, we integrate a wide range of information and introduce two general variations combining graph neural networks with a CNN-based framework. Graph-CNNpred was evaluated for predicting trends in popular stock indices, namely the $\text{S}\&\text{P}$ 500, NASDAQ, DJI, NYSE, and RUSSELL. The results indicate that both versions of Graph-CNNpred significantly outperformed state-of-the-art baseline algorithms. Across all five indices, Graph-CNNpred enhanced prediction performance by approximately $4\%$ to $15\%$in terms of F-measure.

These findings underscore the effectiveness of our approach and underscore the importance of designing CNN structures for tackling stock prediction challenges. While our primary focus was predicting directional movements in stock markets, Graph-CNNpred was successfully integrated into a trading system. This success suggests that further exploration of Graph-CNNpred for trading system applications could open promising research avenues. 

\printbibliography[heading=bibintoc]

\end{document}